**Title:** A New Cosmological Model
**Author:** J.Zzimbe
**Comments:** 18 pages, PDF

It is only human nature to be reluctant to accept something new and different. This feeling is exacerbated given the success of the Big Bang model. However, there are cogent reasons for considering this new model. Some of the advantages of this new model over the existing Big Bang model are as follows: (1) It does not violate any known laws of physics. This contrasts with the Big Bang model where all known laws of physics break down in the consideration of the universe beginning as a singular point of infinite density. (2) The cause behind the initial explosion is provided in this new model. The Big Bang model does not explain what caused the initial explosion. (3) The homogenous nature of the universe is intrinsically explained in this new model. Whereas, the Big Bang model relies on an "extraneous" concept like inflation to explain the homogeneity of the universe. (4) This new model eliminates the mystery of why there is a preponderance of matter over anti-matter. These are some of the advantages of this new model. There are others. Given these qualitative advantages, this new model is worthy of serious consideration even though it is new and different.

The origins of the universe, both in terms of an explanation and what actually transpired, can be broken down into three separate phases. They are as follows.

I.   The starting material of the universe.
II.  The occurrence of "an accident."
III. A series of reactions which culminated in the birth of our universe as we know it.

A full explanation of these three phases is as follows.

<u>I The starting material of the universe.</u>

Prior to the big bang, there was nothing but hydrogen gas. There were no isotopes such as deuterium or tritium, but merely hydrogen. This was the starting material of the universe and the foundation of all subsequent matter that was eventually created. The reader's natural question is, "Where did this hydrogen come from?" Unfortunately, the answer to this question is extraordinarily complex and difficult to grasp.

It is my position that this hydrogen gas was always there. It had no point of origin. To be more precise, the starting material of the universe was infinitely old.* The natural reaction of any person, regardless of whether they are a layman, or a scientist is to claim that such a phenomenon is impossible. Everything had to have some point of origin. However, the reader should keep in mind, that the only reason for such a reaction would be the parameters of human thinking. To be more precise, *everything* that the human mind is cognizant of had some point of origin. Therefore, the natural human reaction to an entity that was always there (infinitely old) is to claim that such a phenomenon is impossible. However, it should be kept in mind that merely because a phenomenon is beyond the realm of human thinking, that doesn't render the occurrence of that phenomenon impossible. By way of superficial analogy, if we insist that it is impossible for hydrogen to have always been there, we become little better than our ancient forefathers who insisted that the earth was the centre of the universe and that everything else revolved around the earth. It was only human arrogance which insisted upon such a position. The ancients expected the universe to "bow down" before man. The reader who claims that everything *must* have a point of origin is little better than those who insisted that the earth was the centre of the universe. The reader may object to this line of argument as the ancients were arguing about the physical proximity of the earth in relation to the rest of the universe. I am referring to a physical process. However, there is no known scientific law which would rule out the possibility of hydrogen gas always being there. All *known* physical laws would allow the starting material of the universe to be infinitely old. Any opposition raised to this possibility would be exclusively predicated upon the parameters of human thinking, and the subsequent insistence that the universe should function in accordance with our thinking.

Obviously, I do not have to remind the reader that we human beings are only an incremental part of this universe. We are in no position to outline how the universe "should behave." We must learn to adapt our thinking to what may be the reality of the universe, even if that reality is completely beyond everything that the human mind knows of, or for that matter can conceptualize. A superficial analogy from the 1920's may serve to elucidate.

If one reads the Feynman Lectures on Physics (volume II, chapter 35), it was the Stern-Gerlach experiment which established the concept of the quantization of angular momentum. Feynman states that the concept came as a substantial shock. Furthermore, he seems to provide some indication that it was a difficult concept to grasp. Intuitively speaking, one may believe that angular momentum could assume any value whatsoever. Prior to the 1920's, we human beings had no frame of reference which would allow any other possibility. (Whether or not the last statement is entirely true, I cannot state as Feynman indicates that the quantization of angular momentum was expected theoretically. However, for the purpose of this paper, we shall assume that there was no frame of reference to allow the conceptualization of the quantization of angular momentum). However, as the Stern-Gerlach experiment has unequivocally shown, angular momentum is indeed quantized. Such a phenomenon comes as a substantial shock since, as has already been stated, the human mind has no frame of reference to enable one to easily grasp such a possibility. Since the quantization of angular momentum is completely beyond the realm of human thinking, should we insist that the Stern-Gerlach experiment is, in some way, invalid? Should we insist that angular momentum can assume any value whatsoever, and that it is impossible for angular momentum to be quantized because of the parameters of human thinking? Of course not! Nature is not going to "bow down" to human thinking. We human beings must learn to adapt our minds to the reality of nature, even if that reality is completely beyond the realm of human thinking.

---

* If the reader encounters no difficulty with this concept, it is fundamentally superfluous to read section I as the rest of section I is exclusively dedicated to expounding upon how this phenomenon is possible despite the fact that it is completely contrary to human thinking. Therefore, please proceed to section II which begins on page four. However, if you insist that it is impossible for something to have always been there, then you must read the rest of section I.

It is no different with the concept of hydrogen gas having no point of origin, but instead always being there. Merely because the human mind has no frame of reference which would allow something to always be there, that doesn't mean it's impossible. The only legitimate justification the reader has for claiming that such a phenomenon is impossible, is if known physical laws are being violated. There are no known physical laws which would prevent hydrogen gas from having no point of origin, but instead being infinitely old.

I can only expound upon the reader's objection (it's impossible for something to have always been there) for so long. In the final analysis, it's a question of learning to adapt one's thinking to this rather difficult concept, despite the fact that there is no frame of reference upon which to rely. A creationist would never be able to grasp this theory as they would be unwilling to implement the requisite effort to grasp the nuances of this concept. They would adamantly insist that everything had to have a point of origin and that it is impossible for something to have always been there. Therefore, this is proof that a supreme deity exists who created the universe. However, I trust scientists would not adopt such an obstinate view, and would make some effort to adapt their minds to (what may be) the reality of nature.

I can summarize "Phase I" of the origins of the universe, (as well as state a precept which will be of assistance in adapting the reader's thinking to the concept of something always being there) under three points. (The precept that will be of assistance is under point 3, indented, and italicized.)

1. It is my position that the starting material of the universe was hydrogen gas. This hydrogen gas had no point of origin, but instead, was always there. It is infinitely old.

2. The natural human reaction to this, is to claim that such a possibility is impossible. Everything must have had some point of origin.

3. There is a reaction of this nature because everything that the human mind knows of, or for that matter can conceptualize, had some point of origin. These are the parameters of human thinking. However, the universe will not be bound by human thinking. Merely because everything that man knows of had a point of origin, that doesn't mean that something could not have always been there. The universe will not "bow down" to mankind. On the contrary, we human beings must learn to modify our thinking to the realities of nature. Or, to put it more cogently,

*The laws of nature do not conform to human thinking.*
*Human thinking must conform to the laws of nature.*

II The occurrence of "an accident"

If I am correct in stating that hydrogen gas was always there, then we can conclude that the birth of the universe was a virtual accident. The full meaning of this can be explained as follows.

Our current estimates of the age of the universe, place the universe's age at approximately twelve billion years (with an error of plus or minus four billion years). Let us take the higher estimate of this and assume the universe is sixteen billion years old. Although this may seem to be an extraordinarily large number, one must realize that sixteen billion years is extraordinarily small *in relation* to hydrogen gas that was always there (infinitely old). It would be appropriate to note that when dealing with a material that is infinitely old, *any* number, regardless of it's magnitude, is quite small. In other words, even if the universe (as we know it) was $10^{1000000}$ years old, even this would be quite young in relation to gas that was always there. Therefore, whatever phenomenon caused the birth of the universe had to be a virtual accident.

It is possible that the reader is somewhat perturbed at labeling the birth of the universe as "an accident." Therefore, it is entirely appropriate to utilize other terminology which conveys the same set of precepts, yet utilizes completely different words. It would be appropriate to state that the birth of the universe was the result of a phenomenon which was *extraordinarily* improbable, yet physically possible. What physical event would be phenomenally improbable, yet physically possible?

My position as to what this event could be is explained via the following equation.

$$\Delta S_{\text{(of a small part of the system)}} < 0 \qquad (1)$$

(Since entropy can potentially refer to various physical phenomena, it must be pointed out that what is being emphasized in this equation is the relationship between order and disorder when dealing with a designated system). In other words, a tiny, incremental portion of the hydrogen gas flowed into an ordered, spherical mass. The reader may immediately object to this position as the laws of thermodynamics seemingly would not allow such a process. However, I should not have to remind the reader that the second law of thermodynamics is not predicated upon the abrogation of physical law. The second law is predicated upon statistical probability. In other words, the second law does not state that a system *cannot* flow from disorder to order. It states that it is phenomenally improbable that a system will flow from disorder to order. The reader may nevertheless object. It is possible to present a scenario which, in the reader's eyes would render my position invalid.

Let us assume we have a box that is 1,000,000 cubic metres in volume. We then place a 1 metre diameter balloon filled with gas in the middle of this box. The gas, *in relation* to the box, is in a state of order. We then burst the balloon and immediately close the box (a closed system). In accordance with the second law of thermodynamics, the gas will spontaneously flow from a state of disorder to order. What is the probability that even a small portion of this gas will spontaneously flow into a small spherical mass that is, let's say, three centimetres in diameter? Obviously, the probability is extremely low. It may be thousands, millions, even billions of years before such a spontaneous flow occurs. Since the probability for such an event within a system that is only 1,000,000 million cubic metres in volume is so low, we can eliminate the possibility of such an event when dealing with a system as large as the universe. If the reader is making a statement of this nature, then you obviously have not read this paper carefully enough. Therefore, I shall repeat myself.

*Whatever* age we place on the universe (as we know it) this age is phenomenally young when compared to the starting material of the universe. The starting material of the universe had no point of origin and is infinitely old. Therefore, we can conclude that whatever was responsible for the birth of the universe was an accident. In other words, to ensure the cohesiveness of this theory, we need an event which was phenomenally improbable, yet physically possible. This is precisely what the second law of thermodynamics provides. Therefore, not only have I *not* opposed the second law, I have worked in precise accordance with it.

When dealing with the origins of the universe, there is more to be stated than merely a small portion of the original gas flowed from a state of disorder to order. Hydrogen has a natural diffusion rate. Let us assume that a small portion of it flowed into a spherical mass that was only one metre in diameter. The (self) gravitational attraction of this mass would be minimal. The diffusion of the hydrogen would exceed the escape velocity of this body, and would flow into a state of disorder. There would then be an inordinate period of time before a small portion of the original gas once again flowed into a state of order. If the spherical mass was too small, the hydrogen would once again flow into disorder. Therefore, a more precise description of what transpired prior to the initial explosion that gave birth to the universe is as follows. First, we must pick an arbitrarily large number for this description. Let us say $10^{1000}$. Every $10^{1000}$ years, a small portion of the system would flow from a state of disorder to order. However, since the escape velocity of a small mass would be minimal, the hydrogen would once again flow away into a state of disorder. This cycle of events would continue (in our example, every $10^{1000}$ years) until enough of the hydrogen flowed into a mass which was large enough to exert a reasonably strong gravitational attraction (large escape velocity) to prevent, over a certain period of time, all of the gas spontaneously diffusing away. Over time, more and more hydrogen atoms would "come within range" of the gravitational attraction of this mass. Naturally, they would be attracted and this mass would grow in size. As it became increasingly larger, the gravitational attraction of this body would dramatically increase. The hydrogen atoms in the universe would be attracted with greater ease. Eventually this mass would grow to *phenomenally large* proportions. I will pick an arbitrary measurement to give some indication as to just how enormous this initial body of hydrogen gas was. In today's terms (dealing with the universe as we know it), this body may have covered an area equivalent to twenty billion galaxies. (The reader will immediately raise an objection to this possibility. It would be the reader's position that at a certain size, without thermonuclear reactions, this body would collapse under self gravitation. I am cognizant of this objection and realize that it constitutes a *very* serious anomaly within this overall theoretical framework. It will, to one extent or another, be dealt with at the end of this paper). When I claim that this mass covered an area equivalent to twenty billion galaxies, I should not be "tied down" to this number. This initial mass may have been smaller or significantly larger. I am merely trying to convey the precept that this initial mass was extraordinarily large.

I have stated that the birth of the universe was the result of an accident. It may be more precise to state that a second accident (or an event which was highly improbable) was also required to give birth to the current universe. This initial mass may have existed for a substantial period of time (millions, or billions of years) before something "happened" to it. The second accident may have been as follows. All the atoms would have had independent motions in the x, y, or z directions. However, at some stage, a sufficient number of these atoms spontaneously moved in the same direction (let's say counter-clockwise) to cause the entire body to start rotating. Once the rotation proceeded at a sufficient speed, all the atoms would be moving in the same direction. Once the atoms stopped moving in arbitray directions, the rotational rate of this initial body would increase.

III A series of reactions which culminated in the birth of our universe as we know it.

Once this body initiated a state of rotation, what would have transpired to the hydrogen atoms? (In the following equation "s" will denote speed).

$$s_{(\text{hydrogen})} = \sqrt{\frac{2GM}{r}} \qquad (2)$$

*If* the speed of the atoms had achieved this particular magnitude, the atoms would have achieved escape velocity and proceeded to flow away from the conglomerate. However, given the enormous mass of the hydrogen conglomerate, it is virtually guaranteed that the following equation would prevail during the initial state of rotation.

$$s_{(\text{hydrogen})} \ll \sqrt{\frac{2GM}{r}} \qquad (3)$$

Consequently, the atoms would have moved closer to the centre and the conglomerate would have condensed.  In other words, upon initiation of a state of rotation, centripetal forces would have started to act and the pressure exerted towards the centre of this body would have increased.  This latter statement may be brought into question as direct result of a dispute over basic science.

If one ties a string to a stone and proceeds to twirl this string about one's head, there will be a centripetal force acting on the stone.  However, the centripetal force does not cause the stone to move any closer to the centre.  It merely keeps the stone rotating in a circle.  This, of course, is valid for a stone being rotated on a string here on earth.  However, when dealing with the origins of the universe, we are not dealing with a stone on a string but instead gas that is being acted upon by gravity.  Since any gas has a "natural freedom of movement", the density of this initial body will increase and the pressure towards the centre will increase.  The second objection (as a result of basic science) may be that the pressure in the centre will be minimal at best. This is because $a = v^2/r$ thereby indicating that acceleration is inversely proportional to radius.  However, this tenet states that if the radius is sufficient in magnitude, it will require a substantial initial force to initiate the rotation.  I have already stated, that a sufficient number of the hydrogen atoms would have spontaneously moved in the same direction in order to initiate the rotation of this body.  In short, the requisite force would have existed in order to initiate rotation.

In order to elucidate on the magnitude of the pressure that would have been exerted towards the centre of this body (even with a *minimal* rate of rotation) two rudimentary physics equations shall be presented.

$$P = \frac{F}{A} \qquad (4)$$

and

$$F = MA \qquad (5)$$

Although we cannot gain a knowledge of the precise quantitative pressure exerted towards the centre (since we are oblivious to the mass or rate of rotation of the hydrogen conglomerate), these two equations will provide a reasonable indication of whether the pressure was minimal or substantial. Firstly, I remind the reader that I have picked an arbitrary measurement of twenty billion galaxies over which the initial spherical mass of hydrogen extended. This entails that the force exerted towards the centre would have been phenomenal as a result of the inordinate amount of mass that such a large conglomerate would inherently entail. (This phenomenon would prevail even if the acceleration of the hydrogen atoms was minimal). This factor, *unto itself*, would have resulted in tremendous pressure exerted towards the centre. However, equation number four dictates that the pressure brought to bear via this phenomenal mass would have been substantially magnified.

Pressure is inversely proportional to area. When dealing with a spherical mass, the further "we proceed" (loosely speaking) towards the centre of the mass, the smaller the physical area. What this obviously entails is that the area in the centre of the hydrogen conglomerate experiences the greatest pressure. Consequently, we have two different factors contributing to the pressure towards the centre. An inordinate mass of material and an increasingly smaller area as we proceed closer towards the centre. What would the combination of these factors have resulted in? *A phenomenally inordinate pressure exerted towards the centre of the hydrogen conglomerate.*

In the presentation of this paper, the science has been fairly straightforward. The difficulties of this paper lie in some of the concepts utilized which are contrary to human thinking and/or scientific training. Obviously it would be difficult to grasp the concept of hydrogen gas always being there since it is completely beyond the realm of human thinking. Furthermore, when I state that a small part of a system spontaneously flows from disorder to order, *a preliminary analysis* may seem to indicate that I have violated the second law of thermodynamics. Unfortunately, I am now forced to introduce a third concept which, initially, may seem to be contrary to existing physical law and would certainly be contrary to a scientist's training.

It is my position that once the rotational rate of this initial mass reached sufficient speed (or to be more precise, the centripetal forces became sufficient in magnitude) *within the core of this mass*, the following transpired. (In the following equation, "X" must be a natural number).

$$(4)(X)_1H^1 \xrightarrow{\text{pressure}} X\,_2He^4 + (2)(X)\boldsymbol{b}^+ + (24.7 \text{ Mev})(X) \qquad (6)$$

Specifically, within the core of this mass, fusion would have been achieved. Any scientist will immediately raise strenuous objections to the position advocated in this equation.

What is specifically being opposed is the word pressure (as opposed to heat). In accordance with your training, it is only very intense heat that will overcome the coulomb repulsion in order to achieve nuclear fusion. However, I am asking the scientist to give proper consideration to my position. Once a centripetal force was exerted on this body, the atoms in the centre would become closer and closer to each other (as established by equations four and five). The coulomb force would indeed keep them apart for a period of time. However, we are fully cognizant of the fact that the coulomb repulsion can be overcome via phenomenally intense forces. Scientists have always been of the opinion that it is only heat that will constitute the requisite phenomenal force to overcome this repulsion. However, if scientists will give due consideration to this, why wouldn't pressure overcome the coulomb repulsion? The atoms would continually get closer and closer to each other. They couldn't move away from each other because of the centripetal force. Does the reader feel the pressures aren't sufficiently intense? Keep in mind that since we are dealing with such a large mass that is focused on a small physical area (the centre of the conglomerate) the pressure towards the centre would be phenomenally intense. As has already been stated, phenomenally intense forces can overcome the coulomb repulsion. If a scientist can overcome his pedantic thinking, I trust that even a modicum of thinking will reveal that it is possible for the core of a *very large* rotating body of hydrogen gas to achieve fusion. (The reader, no doubt, experiences another serious difficulty with a proposal of this nature. Specifically, why is the attainment of fusion being advocated via a centripetal force as opposed to a gravitational collapse? This anomaly, as it pertains to gravity, was briefly mentioned on page six of this paper. Again, it will, to one extent or another, be dealt with at the end of this paper.)

Once this fusion state was initially achieved, the heat produced would expand spherically outwards. It is my position that the initial state of fusion was achieved *exclusively* through pressure. However, any *subsequent* fusion achieved (in relation to the origins of the universe) was the result of a similar process with one additional factor involved (which, for those of you who aren't particularly observant, is stated above the arrow). For the *subsequent* state of fusion, the following equation becomes germane.

$$(4)(X)_1H^1 \xrightarrow{\text{pressure and heat}} X\,_2He^4 + (2)(X)\boldsymbol{b}^+ + (24.7 \text{ Mev})(X) \quad (7)$$

Subsequent to the initial fusion, the combination of the expanding sphere of heat and the existing pressure from the centripetal force would have caused the next "layer" of hydrogen to fuse, thereby producing more helium and more heat. This would result in the next "layer" accomplishing the same thing. This would result in a chain reaction in which more and more hydrogen would have achieved fusion. In essence, a "snowball" effect would have been achieved in which more and more hydrogen was fused with the results of more and more heat and helium being produced.  Would this effect have continued until all of the hydrogen gas had fused?

This snowball effect would not have continued to the point where *all* of the hydrogen of the initial conglomerate had been fused.  There are two reasons for adopting a position of this nature.  Both of them pertain to the geometry of a spherical object and, therefore, to this system (a conglomerate of hydrogen gas with an outward expanding sphere of fusion).  The first reason is explained via the following.

The following equation is directly related to equations four and five

$$P = \frac{1}{r} \qquad (8)$$

Within the parameters of this theory, pressure is inversely proportional to radius for two reasons (which, again, are related to equations four and five).  As we proceed further from the centre, there is less mass being "directed" towards the centre.  Consequently, there is less force.  Secondly, the greater the distance from the centre, the larger the physical area over which fusion would have to be achieved. Since pressure is inversely proportional to area, the pressure would be reduced.  Therefore, equation eight is one reason the snowball effect would not have continued until all of the hydrogen had fused.

As stated previously, there is a second reason why fusion would not have continued to the point where all of the hydrogen fused.  Once again, the "effects" of this particular equation will only become most acute in physical areas that are at a considerable distance from the centre of the conglomerate.

$$_1H^1_{(quantity)} > \Delta Q \qquad (9)$$

In other words, a time would eventually have been reached when the quantity of hydrogen gas exceeded the heat being produced, thereby negating the process of fusion.  As in previous parts of this paper, there may, once again, be objections to this position as a result of alleged scientific inaccuracies.  The reader may be of the view that since there is more hydrogen fusing (as we proceed further from the centre), there is more heat being produced thereby enabling a greater quantity of hydrogen to be fused.  During the initial stages of fusion, the accuracy of this fact is self evident.  However, as the result of the geometry of a sphere, this situation would not have always prevailed.  Eventually, a stage would have been reached where the number of hydrogen atoms *greatly* exceeded  the amount of heat being produced.  Once this stage was reached, there would be insufficient heat to induce fusion within subsequent "layers" of hydrogen.

Consequently, as a result of the geometry of a sphere, the following situation will prevail when the expanding sphere of fusion expands to the point where it is at a considerable distance from the centre. There will be a lack of pressure (as a result of less mass acting over a greater physical area), and an ever increasing number of hydrogen atoms. As a direct result of these two effects, an expanding outward sphere of fusion would eventually cease.

How would this phenomenon culminate in the birth of our universe? This theory must account for the empirical fact that, according to our observations, hydrogen is the most abundant element in the universe. The hydrogen that did not achieve fusion would have been absorbed within the expanding radius of energy. However, how much of the hydrogen was absorbed? Unfortunately, the answer to that question is highly speculative.

In order to answer this question accurately, we must be in possession of *reliable* statistical data. The reader may feel that we are already in possession of such data. However, I feel that such a viewpoint is too simplistic. It would be naive to assume that our state of technology is sufficient to make conclusive observations that can be unequivocally accepted when dealing with a system as large as the universe. A trivial example of this would be the Hubble telescope. Prior to the launching of the Hubble, our observations stated that the universe was comprised of 10 billion galaxies. Subsequent to the Hubble, the number was revised upwards to 50 billion galaxies. The point that I am striving to make is as follows. We cannot be *absolutely certain* that hydrogen is the most abundant element in *all* galaxies. Although our current observations indicate that this is the case, can we be certain that our state of observational technology is sufficient to ensure that this conclusion is correct? What about galaxies that may, potentially, be far beyond the scope of our current observational capabilities. Is hydrogen the most abundant element in these galaxies as well? It would be naive, and simplistic to assume that we have observed the entire universe. It is for this reason that I state the answer to the question (how much of the hydrogen that was not fused was absorbed by the expanding sphere of energy) is highly speculative.

*If* hydrogen is indeed the most abundant element in *all* galaxies, then there are three possibilities in deciding how much of the original hydrogen gas was absorbed by the expanding sphere of energy.

(1) The hydrogen that was not fused may have been just barely sufficient to be absorbed by the energy as it expanded outwards so that none of the original hydrogen was left. In other words, *all* of the unfused hydrogen was absorbed. In my opinion, the probability of this is extremely low.
(2) The hydrogen that was not fused exceeded (by a small amount) the amount of energy that was expanding outwards. This would entail that a small amount of hydrogen would be left once the sphere of energy had expanded beyond it. The self gravitation of such a body would be minimal and the hydrogen would diffuse into the vacuum of outer space.
(3) The hydrogen that was not fused exceeded, to a very large extent, the expanding sphere of energy. Since the self gravitation of this body would be large, it would have been absorbed into the original core of hydrogen that was *first* fused thereby dramatically increasing its mass.

In my opinion, the third possibility would be the most probable scenario for the following reason. The amount of energy created during the birth of the universe would have been absolutely phenomenal. The energy would have been so intense, that it would have been sufficient to "blast away" any of the remaining hydrogen that was unfused. In other words, the physical proximity of the unfused hydrogen gas would have been blown away to an area far away from the expanding sphere of energy as a result of the intensity of the initial blast. This would have prevented hydrogen from being the most abundant element in the observed galaxies. However, since hydrogen *is* the most abundant element (in accordance with our observations) it could not have been blasted away once the hydrogen stopped fusing. For the hydrogen to remain where it was, (and thereby be absorbed) the mass of the unfused hydrogen must have been quite substantial.

In this analysis, I have proceeded from the position that hydrogen is the most abundant element in all galaxies. However, what if hydrogen was not the most abundant element in all galaxies? It is possible that the hydrogen not fused was significantly less than the expanding sphere of energy. What scenario would be presented then? Let us analyze the hydrogen that was fused. The fused hydrogen that was furthest away from the centre of the *initial* conglomerate of hydrogen gas would have been the last to fuse and thereby establish energy and helium. Therefore, this energy would have been the first to absorb the hydrogen that was not fused. If the hydrogen not fused was minimal (relatively speaking), then the energy/helium first established (closest to the centre) would not have absorbed any hydrogen by the time it expanded sufficiently outwards. This, of course, would be for the simple reason that there was no hydrogen left to absorb as it was already absorbed by the energy/helium that was further away from the centre. Under these circumstances it would be possible to establish the following scenario. There may be galaxies which are helium rich because of the initial fusion that took place and the lack of hydrogen to absorb. Current models of stellar evolution would dictate that such stars would quickly reach one of the final stages of stellar evolution depending upon the mass of any initial stars formed. In other words, it is possible that there are galaxies which are completely devoid of main sequence stars, but are, instead, replete with white dwarfs, neutron stars, and black holes. If such galaxies existed, our current observational capabilities would be unable to detect them. However, as has already been stated, we must wait until we reach the technological stage where we can be absolutely certain as to the reliability of our empirical observations before drawing such conclusions.

We can now return to a more tangible theory of the origins of the universe that does not rely so heavily on speculation as a result of the lack of *conclusive* empirical data. The expanding sphere of energy would have resulted in enormous vortices of energy which contained substantial quantities of hydrogen and helium. In accordance with our observations of an expanding universe, these vortices of energy would have moved further away from each other as a result of their intrinsic energy *as well* as the hydrogen and helium within the vortex which would have provided mass and subsequent momentum. Therefore, we have individual vortices of energy comprised of hydrogen and helium moving away from each other. Eventually, each of these vortices would have formed a singular galaxy upon cooling. Obviously, these vortices would have been inordinately large. However, when dealing with the cooling stages of these vortices, there is more to be said. These vortices would have to lose heat to the vacuum of outer space in order to achieve the requisite cooling. However, in the beginning of the expansion, they would have been reasonably close to each other. As one vortex lost some heat, another vortex may have been "close behind" (loosely speaking) and absorbed the heat that the first vortex lost.  This would circumvent the cooling effect needed to give birth to a galaxy.  Therefore, before galaxy formation transpired, the vortices had to be sufficiently far apart from each other in order to lose their heat to the vacuum of outer space.  As the vortices became increasingly far apart from each other, they continued to lose heat and eventually became cool enough so that the vortex could "break up" to form individual stars.  The term "break up" is utilized in extremely loose terms.  Obviously, the vortex would have stayed together in order to establish a galaxy.  However, once it lost a sufficient amount of heat, it would not remain as a cohesive vortex of energy, but would instead result in the formation of individual stars within the system.

An analysis would seemingly indicate the following. If the vortex were spiral in nature, a spiral galaxy would be established. If the vortex (at the time of galaxy formation) was thicker at the centre than at the edges, an elliptic galaxy would form.  If the initial vortex had no recognizable shape, an irregular galaxy would be established.

The precept of vortices of energy losing heat to the vacuum of outer space raises an interesting issue.  Just how old is the universe?  In deciding the age of the universe, we have always attempted to ascertain the amount of matter in the universe and what bearing this would have on the expansion rate of the universe.  However, this theory presents new parameters to take into consideration.  To briefly recapitulate, a vortex of energy would have to lose heat to the vacuum of outer space before it became cool enough to form a galaxy.  This means that at the beginning of the expansion, each vortex possessed a tremendous amount of energy, and subsequently, a very high rate of speed.  Gravity would not have been able to "counter" this rate of expansion.  In other words, in the beginning, the rate of expansion would have been *significantly* higher than it is today.  As time went on, the vortices would have lost energy, and slowed.  This process would continue until galaxy formation was achieved and the current rate of expansion would have been established.  In other words, the rate of expansion was very high in the beginning, but gradually slowed as a result of  vortices losing energy to the vacuum of outer space. This is another factor to take into account when estimating the current age of the universe.  It is my opinion that the universe is much younger than current estimates would indicate.

There is another aspect to deal with which this theory raises.  When dealing with the question of how much unfused hydrogen was absorbed by the expanding vortex of energy, I outlined three possibilities.  Under possibility number three, I stated that if the unfused hydrogen was sufficient in mass, it would have collapsed into the core that was *first* fused.  This core would have existed regardless of whether hydrogen gas collapsed into it or not.  It would have been inherently different from the vortices of energy that were subsequently created.  Firstly, it would have been *much* more massive.  Secondly, it would not have been moving in the same manner that the other vortices of energy were moving.  The vortices subsequently created were moving *away from this central area*.  This central area would not have been moving away from anything.  The only (probable) movement it experienced, would have been rotational in nature.  This raises an interesting question as to what might be at the centre of the universe.  Current models dictate that by this time, this core would have evolved into a massive black hole since it was the very first thing created as a result of fusion.  However, future theoretical models may dramatically alter this possibility.

Thus far, this theory has predominantly dealt with the formation of the universe as a whole, and other large scale structures such as galaxies. However, this theory also illuminates two other mysteries pertaining to the universe. Specifically, the formation of binary stars and the paradox pertaining to globular clusters (why they are seemingly older than the universe itself). Firstly, we will deal with binary stars.

In order to ensure that the explanation behind the formation of binary stars is lucid, we will first review some first year physics as it pertains to centrifugal and centripetal forces. In a previous part of this paper, an analogy pertaining to a centripetal force was drawn via a stone on a string. In order to explain the formation of binary stars, we will maintain that analogy. However, the string will be replaced with an elastic band. Or, to be more specific, something that can be easily deformed. If one were to swing this elastic with the stone at the end of it around one's head, what will transpire as the speed increases? By virtue of Newton's third law, there will be a centrifugal force present "in response" to the centripetal force being exerted on the stone. However, because there is a mass at the end, the elastic will become longer since the stone is now pulling on the elastic. If the speed of the rotation continues to increase in such a manner as to not only compensate for the increased radius (when dealing with centripetal/centrifugal forces radius and the force required to induce rotation are inversely proportional), but also increase the speed of the rotation, the deformation of the elastic will eventually become substantial enough to cause the elastic to break (assuming the elastic is not inordinately strong). To put it in slightly more succinct terms, the two factors which will induce this effect are the existence of a substance which can be easily "deformed" (speaking loosely), and a mass at the end of that substance. This provides the background information to comprehend the formation of binary stars within galaxies in the early universe.

When an individual galaxy began to cool sufficiently, stellar birth would have been initiated when vortices of energy *within* a designated galaxy began to lose their heat and break up into individual vortices of energy within the galaxy. Some of these vortices would have resulted in binary stars via the following mechanism. An individual vortex would have contained hydrogen atoms and, possibly, some helium atoms. Since there was nothing to "attach" these atoms to the centre of the vortex, they would have been "thrown" to the outer periphery of the vortex during the course of its rotation. At this stage, this vortex would be somewhat comparable to the stone on an elastic. Namely, there would be mass at the outer edge, and the energy could potentially break up. As the rotation of the vortex continued, the mass at the edge would have induced a sufficient centrifugal force to cause the energy towards the centre to "break apart". At this stage, a binary star would be formed. This could only have transpired when stellar birth within a galaxy was being initiated for the following reason. If the energy within a galaxy did not begin to break up into individual vortices, the process of binary formation would have been impossible. A vortex which had the "potential" to become a binary could not become a binary unless it had a sufficient amount of "room" (loosely speaking) to do so. In other words, if a vortex was *completely surrounded* by energy, it would not be able to separate into two components. However, if there was "empty space" (a vacuum) around it, then there would be sufficient "room" to enable the separation of the vortex into two vortices.

Some may feel that this model of binary formation is flawed. For a vortex to divide into (what will eventually become) two individual stars, the hydrogen and helium atoms would have to be concentrated at two opposite "ends" of a vortex. In all probability, the atoms would have been equally distributed on the outer periphery of the vortex. Therefore, this model of binary formation cannot be correct. I would not view this precept in the capacity of a flaw which would render this theory incorrect. It would be more in the capacity of a question which must be more precisely answered in the future. The model of binary formation just outlined is fundamentally sound. When it comes to binary formation, why there would *not* be an equitable distribution of atoms on the outer periphery is simply an unanswered question. It may simply be the spatial configuration of the vortex which, somehow, causes the atoms to conglomerate at two opposite ends of the vortex. Or, for some reason, there may be some factor which causes atoms to *suddenly* be "thrown" to two opposite ends of the vortex. Again, this is merely an unanswered question, not necessarily a flaw which invalidates the theory.

This new cosmological model also illuminates the mystery pertaining to globular clusters. Specifically, why are they seemingly older than the universe itself? *In relation to galaxy formation*, globular clusters were the first entities to be formed. (This doesn't mean that globular clusters were the very first entities in the universe to form. The very first complete entity to be formed was at the core of the conglomerate of hydrogen in which fusion was first achieved. Globular clusters were only the first entities *within galaxies* to form). The analysis behind this is as follows. As has previously been stated in this paper, for a large vortex of energy to become a galaxy, a certain "cooling down" effect would have to prevail. Once various vortices were a sufficient distance from each other, this effect would be initiated. Once it was initiated, where would the "cooling down" effect be greatest? Obviously on the outer periphery of the vortex. The rationale behind this is as follows. When considering the outer periphery, there is nothing but the vacuum of outer space immediately surrounding it. Consequently, heat will easily dissipate from the outer periphery to this vacuum. However, when considering the innermost areas of the vortex, these inner areas are *not* surrounded by a vacuum. They are surrounded by more energy which is, obviously, hot. Therefore, it would take much longer for these inner areas to cool down and initiate stellar formation. However, the individual vortices on the outer periphery of the overall vortex (the one which will culminate in the establishment of a galaxy) would have lost their heat far more quickly. Consequently, they would have established themselves as globular clusters well in advance to the rest of the vortex becoming individual stars. Since globular clusters were the earliest parts of galaxies to form, this would explain why they are seemingly older than the universe itself.

The reader may not accept this model. From all outward appearances, this model is in contradiction to observations. This model is adopting the position that globulars dominated the outer periphery of a galaxy. Although there are some globulars on the outer periphery of galaxies, observations show that they tend to dominate within the central areas of galaxies. This model of globulars can still be maintained as accurate via the following explanation. When globulars first formed, they could not move from the outer periphery towards the central area (as a result of a centripetal force) since the energy of the vortex would have physically prevented such movement. They would have only moved to the centre (via a centripetal force) when the energy of the vortex dissipated, stars began to form, and there was a sufficient amount of empty space among the various (newly formed) stars. Once this empty space was created, there would have been nothing to impede their movement, and most of them would have moved towards the centre because of the centripetal force experienced via the rotation of the galaxy. (Obviously all of them didn't move and there must have been physical impediments to hinder their movement).

      Therefore, the "mystery" behind globular clusters is answered and a (fairly) complete model of binary formation is provided.

I will now deal with potential opposition to this theory.

      The first area for opposition may lie in the theoretical possibility of proton decay. In other words, it is impossible for the starting material of the universe (hydrogen gas) to be infinitely old as the protons would have eventually decayed. As of this writing, these are grossly insufficient grounds for objection. This theory of the origins of the universe has worked within the parameters of known physical laws. Proton decay is a phenomenon which is purely theoretical. It is yet to be experimentally verified. Therefore, until proton decay is experimentally verified, the theoretical possibility of proton decay is insufficient to raise appropriate grounds for opposition.

      The second area for opposition would pertain to the current models of stellar formation. Firstly, once this initial conglomerate of gas reached a certain mass, self gravitation would have prevailed and it would have gravitationally collapsed *before* any centripetal forces could have induced fusion within the core. Furthermore, even if this mass did not collapse under self gravitation, and instead was able to initiate fusion in order to form a huge star, it would have eventually established an enormous supernova type II. When this supernova exploded, the subsequent matter created would not explain the current homogenous nature of the universe.

      In science, theories are not written in stone. Theories are subject to change, and experimental facts are open to new theoretical interpretations. This is particularly acute in astronomy as we cannot conduct controlled experiments as we can in physics. Nobody has actually *seen* a star forming. We've made theoretical inferences based upon experimental observations. It wouldn't be appropriate to dismiss this theory on the grounds that it is in opposition to the current models of stellar formation for the following reasons.

      Firstly, this theory conforms to known observations pertaining to the universe. Secondly, and far more importantly, this theory offers significant advantages over the existing Big Bang model. Quantitatively speaking, there are a few advantages that this model offers over the existing Big Bang model. However, *qualitatively* speaking these advantages are quite significant. Let us now compare and contrast the Big Bang model with this model of the origins of the universe.

| The Big Bang Model | This Model |
|---|---|
| 1. There is *absolutely* no explanation as to what transpired prior to the big bang. | There is a full explanation as to *exactly* what transpired prior to the big bang. |
| 2. The initial mass of the universe was confined to a singular point of infinite density. Because of this all known laws of physics break down. | In this model, no known laws of physics are violated, not even the second law of thermodynamics. |
| 3. It is completely unknown why the initial explosion occurred. | The cause behind the initial explosion is provided. |
| 4. The most common explanation of why the universe is homogenous is to resort to an "extraneous" concept like inflation. | The homogenous nature of the universe is explained without resorting to "extraneous" concepts like inflation, as the universe originally arose from a homogenous body (a large spherical conglomerate of hydrogen gas). |
| 5. It is unknown why there is a preponderance of matter over anti-matter. | The starting material of the universe was matter and therefore there is a preponderance of it. In the same way that anti-matter is created in the high energy systems of colliders, anti-matter was created during the initial explosion. |

The reader should be able to see that when the two models of the origins of the universe are compared and contrasted, this model provides significant advantages over the existing Big Bang model. Therefore, when we have a theory that conforms to known observational evidence, and provides the magnitude of qualitative advantages provided by this theory, it is not legitimate to dismiss it on the sole grounds that it opposes current theoretical models of stellar formation.

The final area of opposition has already been mentioned (briefly) and is quite significant. Namely, we cannot have a conglomerate of hydrogen gas that encompasses a physical area equivalent to twenty billion galaxies as there would be a gravitational collapse *well* before it reached proportions of this magnitude. Although this point of opposition can be dealt with in full, there are certain nuances and difficulties in effectively dealing with it within the parameters of this paper. An explanation is as follows.

Complete papers have been written to explain why the attainment of fusion via a centripetal force is being advocated as opposed to a gravitational collapse. The reader's natural question is, if these papers have been written why not post them? Firstly, the material is quite lengthy (approximately fifty pages). Secondly (and far more importantly) the theoretical work is very unorthodox and proposes radical new concepts. I believe that the majority of physicists/astronomers would be opposed to work that falls within these parameters (even if the work promises to be an advancement for science). Given the fact that scientists are busy people with their own research, teaching, and lives outside of science it simply would not be fair to post material of this nature (namely, lengthy and unorthodox).  As it stands, some of the concepts proposed in *this* paper are very different and far removed from the mainstream.  I am currently waiting to see the nature of the referee's comments from the journal that this paper has been submitted to prior to proceeding with the dissemination of my work pertaining to gravity.  However, since the material pertaining to gravity has been written, it can, and will, be provided upon request.  However, they will only be provided to scientists who know *in advance* that the material is as "voluminous" as it is, and (most importantly of all) *very* unorthodox.  (In other words, it would not be fair to "ambush" readers of this archive with material of this nature.)  Please send an E-mail with the subject line Request the Gravitational Theory and the paper will be provided.